\documentclass[
 aip,
 pof,
 superscriptaddress,
 article-title,
 amsmath,amssymb,
 preprint
]{revtex4-1}

\usepackage{graphicx}
\usepackage{dcolumn}
\usepackage{bm}
\usepackage[utf8]{inputenc}
\usepackage[T1]{fontenc}
\usepackage{CJKutf8}
\usepackage{mathptmx}
\usepackage{etoolbox}

\makeatletter
\def\@email#1#2{%
 \endgroup
 \patchcmd{\titleblock@produce}
  {\frontmatter@RRAPformat}
  {\frontmatter@RRAPformat{\produce@RRAP{*#1\href{mailto:#2}{#2}}}\frontmatter@RRAPformat}
  {}{}
}%
\makeatother

\begin{document}
\raggedbottom
\clubpenalty=10000
\widowpenalty=10000
\brokenpenalty=10000

\title{Generation and characterization of unsteady inflows in a fan--vane wind tunnel}

\author{Yutong Liu}
\email{yutongliu@stu.hit.edu.cn}
\affiliation{School of Intelligence Science and Engineering, Harbin Institute of Technology, Shenzhen 518055, People’s Republic of China}

\author{Gang Hu}
\affiliation{School of Intelligent Civil and Ocean Engineering, Harbin Institute of Technology, Shenzhen 518055, People’s Republic of China}
\affiliation{Institute of Low-Altitude Science and Technology, Harbin Institute of Technology, Shenzhen 518055, People’s Republic of China}

\author{Hongwei Zhang}
\affiliation{School of Intelligence Science and Engineering, Harbin Institute of Technology, Shenzhen 518055, People’s Republic of China}

\date{\today}

\begin{abstract}
Wind-tunnel experiments often require inflows with specific combinations of turbulence intensity and integral length scale because both can influence the measured response.
To broaden the range of attainable inflows, this study uses pseudorandom vane excitation to generate large-scale turbulence and examines how both quantities vary with the amplitude and spectral shape of the vane-angle command.
In each test, the 15 servo-driven vanes downstream of a 768-fan array received the same angle command, and all fans operated at a common, constant duty cycle.
Hot-wire measurements covered twenty heights and selected spanwise and downstream positions.
At a fan-control duty cycle of 60\%, increasing vane angular amplitude raised profile-mean turbulence intensity from 7.3\% to 31.8\% under periodic forcing and from 6.2\% to 23.0\% under pseudorandom forcing.
With the normalized pseudorandom command spectrum held fixed, larger angular amplitudes increased the low-frequency share of velocity variance, accompanied by longer correlation times and larger integral scales.
At a root-mean-square angular amplitude of $15^\circ$, changing the command spectrum yielded profile-mean integral length scales of 0.88--1.67 m at similar turbulence intensities.
The principal features of the velocity spectra persisted between 2 and 4 m downstream, while integral scales continued to change.
Both forcing types exhibited middle-region minima in turbulence intensity and integral scale.
Regional analysis of a representative pseudorandom case linked these minima to weaker low-frequency fluctuations and shorter correlation times.
These findings provide a basis for selecting vane inputs according to the turbulence-intensity and integral-scale requirements of the intended test region.
\end{abstract}

\maketitle

\section*{Nomenclature}
\noindent\begin{tabular}{@{}ll@{}}
$A_p, A_r$ & Peak and RMS vane-command amplitudes (deg) \\
$CV_U$ & Coefficient of variation of the mean-velocity profile (\%) \\
$f, f_0, f_s$ & Frequency, forcing frequency and sampling frequency (Hz) \\
$h_f$ & Vertical fan-row pitch (m) \\
$H, W$ & Clear height and width of the test section (m) \\
$I_u$ & Streamwise fluctuation intensity, $\sigma_u/U$ (reported in \%) \\
$L_u$ & Streamwise integral length scale, $UT_u$ (m) \\
$S_{uu}$ & One-sided streamwise velocity PSD (m$^2$/s) \\
$T_u, T_\theta$ & Velocity and command integral times to the first zero crossing (s) \\
$u, U, U_d$ & Instantaneous, local mean and design velocities (m/s) \\
$x,y,z$ & Streamwise, transverse and vertical coordinates (m) \\
\multicolumn{2}{@{}l}{\textit{Greek symbols and notation}} \\
$\theta_j$ & Commanded angle of vane $j$ (deg) \\
$\lambda$ & Input-spectrum length parameter (m) \\
$\rho_{uu},\rho_{\theta\theta}$ & Normalized velocity and command autocorrelations \\
$\sigma_u$ & Sample standard deviation of streamwise velocity (m/s) \\
$\tau$ & Time lag (s) \\
$\overline{(\cdot)}$ & Equal-weight mean over the measured heights \\
\multicolumn{2}{@{}l}{\textit{Abbreviations and input labels}} \\
ACF, PSD & Autocorrelation function; power spectral density \\
MAE & Mean absolute error \\
PWM, RMS & Pulse-width modulation; root mean square \\
pp & Percentage points \\
Periodic & Sinusoidal vane forcing \\
LU05, LU10, LU20 & Prescribed spectra with $\lambda=0.5$, 1 and 2 m \\
\end{tabular}

\section{Introduction}
\label{sec:introduction}

Controlled unsteady inflows support experiments on building wind loads, wind-energy conversion, gust response and rejection in unmanned aerial vehicles, and convective heat transfer.\cite{mokhtar_investigating_2025,schottler_impact_2017,kubo_preliminary_2016,haughn_deep_2024,sak_role_2007}
These applications require more than a prescribed mean velocity. Fluctuation intensity describes the overall fluctuation level, whereas the spectrum and integral scale describe how those fluctuations are distributed in frequency and correlated in time or space.
For example, incoming integral scale can affect roof-pressure correlations and peak suction at similar turbulence intensity.\cite{mokhtar_investigating_2025}
Experiments on a heated cylinder likewise show sensitivity to both turbulence intensity and integral scale.\cite{sak_role_2007}
An inflow generator should therefore be characterized through the flow delivered at the test location, including changes in other statistics when one input is adjusted.

Several complementary approaches are available for producing such flows. Spires and roughness elements provide established means of shaping boundary-layer inflows.\cite{counihan_improved_1969,cermak_wind_1981,irwin_spires_1981}
Multi-fan wind tunnels distribute the driving inputs across an inlet, enabling prescribed mean-velocity profiles and time-dependent velocity histories.\cite{nishi_turbulence_1997,kikitsu_flow_1999,cao_reproduction_2002,smith_simplified_2012}
More recent multi-fan systems have expanded modular construction and operating ranges,\cite{kikuchi_low-cost_2019,veismann_low_density_2021,liu_aerodynamic_2025} while directional and latticed operation demonstrate how input arrangement affects the delivered flow.\cite{wang_multiple-fan_2018,wang_latticed_2019}
Cao et al.\cite{cao_actively_2001} combined fans and airfoils to reproduce boundary-layer statistics, illustrating the value of using different actuator types together.

Moving grids and vanes offer a direct way to introduce disturbances into an established mean flow.\cite{makita_active_1991,mydlarski_turbulent_2017,shet_optimizing_2020}
Their driving protocols can vary fluctuation intensity, correlations and anisotropy,\cite{cekli_tailoring_2010,griffin_control_2019,cekli_stirring_2020} and can generate intermittent inflows with large correlation scales.\cite{knebel_atmospheric_2011}
Hybrid facilities combine active modulation and adjustable roughness to broaden the attainable profiles and low-frequency content.\cite{mokhtar_automated_2024,pinyochotiwong_advancements_2025}
Coordinated fan and airfoil-fin inputs have also been used to generate two-dimensional non-synoptic flows.\cite{zhao_reproduction_2024}
Oscillating-vane experiments have characterized gust amplitude, frequency and downstream development,\cite{french_design_2021} related longitudinal modulation to blockage,\cite{wester_how_2022} and identified frequency-dependent limits on the dynamic response.\cite{neuhaus_exploring_2021}

The test facility incorporates fifteen servo-driven flat-plate vanes immediately downstream of a 768-fan array based on the modular architecture of Liu et al.\cite{liu_aerodynamic_2025}
However, the different acceleration and deceleration responses of these fans complicate rapid velocity modulation through fan rotational speed control alone.
Previous passive-spire tests in this facility yielded centimetre-scale integral lengths,\cite{hong_design_2024} motivating an approach that can increase turbulence scale as well as fluctuation intensity.
Vane motion provides an additional means of generating fluctuations by changing flow blockage and deflection, allowing the disturbance input to be varied while the fan commands remain constant.

Selecting vane inputs requires understanding how angular amplitude and the prescribed command spectrum affect both fluctuation intensity and integral scale.
Previous studies have shown that input amplitude can affect both quantities in multi-fan forcing,\cite{ozono_realization_2018} and that integral scales can continue to develop downstream under fixed active-grid driving conditions.\cite{zheng_turbulent_2021}
For the present fan--vane system, this motivates examining amplitude effects at a fixed normalized command spectrum and spectral effects at a fixed root-mean-square angular amplitude.
A further question is how the resulting intensity and scale vary with position under common vane commands.

To address these questions, this study evaluates prescribed-spectrum pseudorandom vane excitation for generating large-scale turbulence, using fixed-vane and periodic cases for comparison.
All vanes receive the same command.
Twenty-height profiles quantify changes in mean velocity and turbulence intensity.
Spectra and autocorrelations are used to examine the scale changes associated with angular amplitude and the differences between upper, middle and lower regions.
Comparisons across fan PWM settings and spanwise and downstream positions assess the spatial and operating dependence of the measured response.
The experiments and analysis are described in Sec.~\ref{sec:experimental_methods}, the results and discussion in Sec.~\ref{sec:results}, and the conclusions in Sec.~\ref{sec:conclusions}.
\section{Experimental methods}
\label{sec:experimental_methods}

\subsection{Experimental facility}
\label{subsec:p1-facility}
\begin{figure*}[!htbp]
\centering
\includegraphics[width=\linewidth]{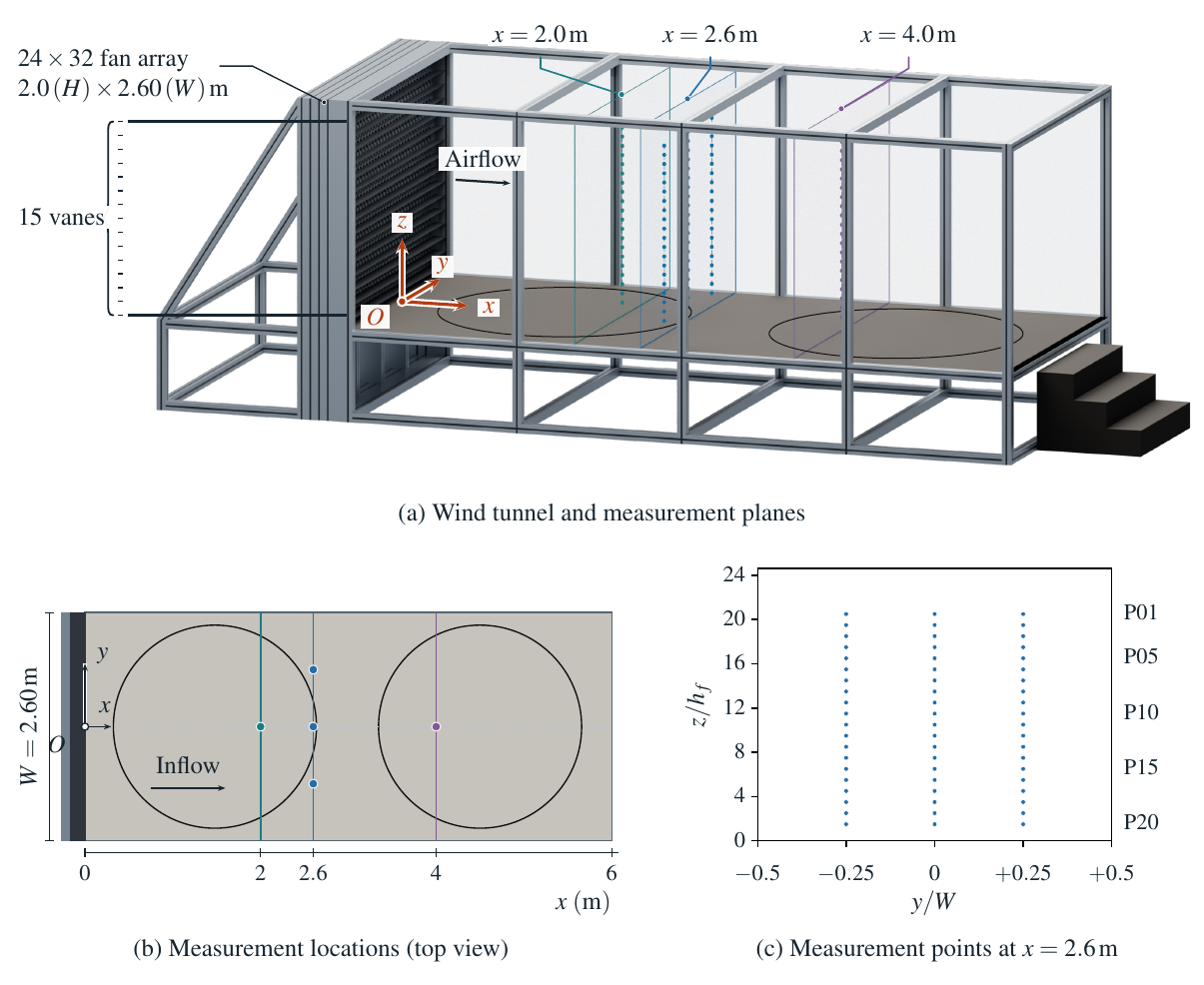}
\caption{Wind tunnel and measurement layout: (a) 24 rows of 32 fans, 15 vanes and three measurement planes; (b) five vertical traverse locations, viewed from above; (c) twenty measurement heights at each of three spanwise positions at $x=2.6$ m. Here, $H$ and $W$ denote the clear test-section height and width.}
\label{fig:facility}
\end{figure*}

\begin{figure*}[!htbp]
\centering
\includegraphics[width=\linewidth]{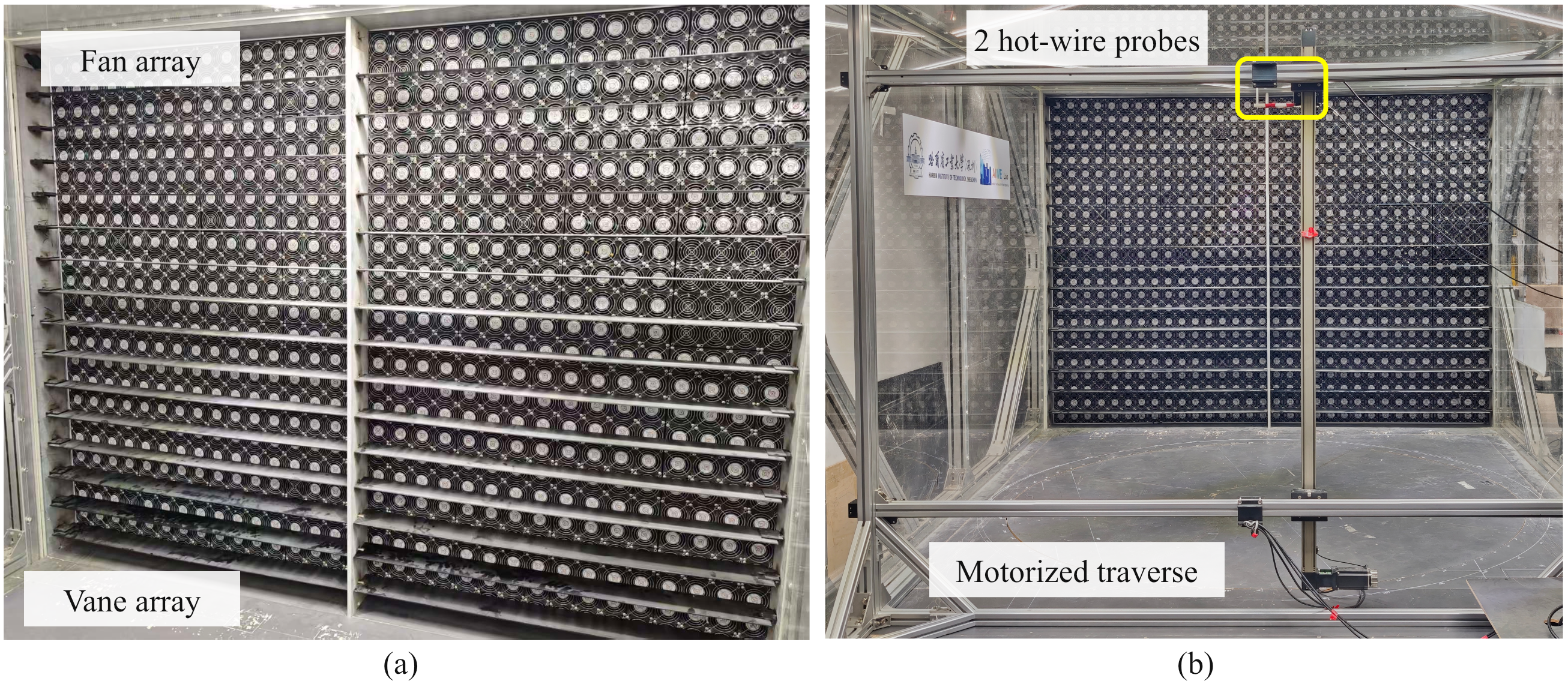}
\caption{Photographs of the facility: (a) fan and vane arrays; (b) view from downstream showing the motorized traverse and two hot-wire probes.}
\label{fig:photographs}
\end{figure*}

The experiments were conducted in the multi-fan wind tunnel at Harbin Institute of Technology, Shenzhen.
Its inlet contains 768 fans arranged in 24 rows and 32 columns, followed by fifteen vanes and a 6.0 m test section (Figs.~\ref{fig:facility} and \ref{fig:photographs}).
The clear width and height are $W=2.60$ m and $H=2.0$ m, respectively.
The nominal vertical fan-row pitch is $h_f=81.25$ mm, including the assembly gaps between the fan units.
The system uses the AVC DFPA0880B8U contra-rotating axial fans and PWM speed-control technology described by Liu et al.\cite{liu_aerodynamic_2025}
Each fan measures $80\times80\times80$ mm and can be controlled individually through its PWM input.
In this study, all fans operated at the same constant duty cycle within each experiment.
Changing this common setting provided different mean-flow operating points for the vane-response comparisons.

The array consists of fifteen flat-plate vanes immediately downstream of the fan outlet.
Each vane spans 2.58 m, extending across nearly the full test-section width, and is driven by a separate servo about a spanwise axis.
The vanes are numbered V1--V15 from top to bottom; their chord lengths are 80 mm for V1--V6, 100 mm for V7--V12 and 120 mm for V13--V15.
The axes are spaced 122 mm apart and lie at the middle of the 0.18 m deep vane module.
At zero angle the plates are horizontal and aligned with the mean flow.
Rotation changes the area presented to the approaching flow and its deflection through the vane array.
All fifteen vanes received the same angular waveform and amplitude and moved synchronously during a given experiment.
This common input allows the spatial response to be examined without imposing different commands at different heights.
The test section has a wooden floor and transparent side and roof panels, enclosing the flow between the generator and the downstream outlet.
Principal dimensions and the measurement layout are summarized in Table~\ref{tab:facility}.

\begin{table}[!htbp]
\caption{Principal facility and measurement parameters.}
\label{tab:facility}
\begin{ruledtabular}
\begin{tabular}{ll}
Parameter & Specification \\
\hline
\multicolumn{2}{l}{\textit{Fan system and test section}} \\
Fan arrangement (rows $\times$ columns) & $24\times32$ (768 fans) \\
Fan-unit dimensions & $80\times80\times80$ mm \\
Vertical fan-row pitch, $h_f$ & 81.25 mm \\
Test section (length $\times$ width $\times$ height) & $6.0\times2.60\times2.0$ m \\
\multicolumn{2}{l}{\textit{Vane array}} \\
Vane configuration & 15 servo-driven flat plates \\
Vertical spacing between vane axes & 122 mm \\
Chord: V1--V6 / V7--V12 / V13--V15 & 80 / 100 / 120 mm \\
Vane span & 2.58 m \\
\multicolumn{2}{l}{\textit{Measurement layout}} \\
Velocity sensor & Single-wire hot-wire probe \\
Heights per profile & 20; fan rows 4--23 from the top \\
Height range, $z/h_f$ & 1.5--20.5 \\
Streamwise positions, $x$ & 2.0, 2.6, 4.0 m \\
Spanwise positions, $y/W$ (at $x=2.6$ m) & $-0.25$, 0, $+0.25$ \\
\end{tabular}
\end{ruledtabular}
\end{table}

\subsection{Velocity measurements and flow statistics}
\label{subsec:measurements}
Streamwise velocity was measured at 2048 Hz using two single-wire probes connected to a Hanghua CTA04-EDU constant-temperature anemometer.
Each probe was calibrated at eight reference velocities over 0--14.5 m/s using a dedicated velocity calibrator; a fourth-order polynomial related its output voltage to velocity.
Hot-wire recalibration was performed during the measurement campaign.
A platinum temperature probe connected to the anemometer provided temperature compensation during each round of measurement.
The probes were mounted on a stepper-motor traverse with a vertical separation of one fan-row pitch.
At each of ten traverse positions, velocity was recorded simultaneously at two adjacent heights.
Local statistics from these measurements were combined into a twenty-height profile.

The origin is at the downstream face of the vane module on the floor and centerline. The coordinates $x$, $y$ and $z$ point downstream, across the width and upward, respectively.
The hot wires were positioned at the center heights of fan rows 4--23, numbered from the top of the array.
These twenty heights are spaced by $h_f$ and span $1.5 \le z/h_f \le 20.5$, corresponding to approximately 0.122--1.666 m above the floor.
Most comparisons were made at $x=2.6$ m and $y/W=0$. Additional profiles were acquired at $y/W=\pm0.25$ in that plane and on the centerline at $x=2.0$ and 4.0 m.
The off-center positions are therefore $y=\pm0.65$ m.
These heights lie outside the viscous region and narrow wall boundary layers, although the enclosure and measurement support can influence the spatial response.

Velocity was recorded for 20 or 30 s at each traverse position.
The active-input spectral comparisons use 30 s records; for pseudorandom excitation, this duration covers one command period.
The fan duty cycle and vane excitation parameters remained unchanged throughout acquisition of each profile, with the prescribed vane motion running continuously for active inputs.

The dataset comprises 41 profile records covering 35 combinations of input and measurement position.
Six combinations were measured twice to assess the repeatability of the measured flow statistics; the paired differences are reported in Appendix~\ref{app:repeatability}.
For repeated measurements, flow statistics were calculated separately for each record and then averaged for comparison.

For a record of duration $T$, the velocity is decomposed into its mean and fluctuation,
\begin{equation}
u(t)=U+u'(t),\qquad U=\frac{1}{T}\int_0^T u(t)\,\mathrm{d}t,
\qquad I_u=\frac{\sigma_u}{U},
\label{eq:velocity}
\end{equation}
where $\sigma_u$ is the sample standard deviation of the streamwise velocity fluctuation $u'(t)$.
$I_u$ is reported as a percentage. For periodic forcing it includes the imposed organized fluctuation as well as the fluctuations superposed on it.
An overbar on a local statistic denotes an equal-weight average over the $N_p=20$ measured heights. The nonuniformity of the mean-velocity profile is summarized by
\begin{equation}
\overline U=\frac{1}{N_p}\sum_{i=1}^{N_p}U_i,\qquad
CV_U=\frac{100\%}{\overline U}
\left[\frac{1}{N_p-1}\sum_{i=1}^{N_p}(U_i-\overline U)^2\right]^{1/2}.
\label{eq:profile}
\end{equation}
$CV_U$ describes spatial variation along the sampled profile, while $I_u$ describes temporal variation at a point.
Profile means of $I_u$, $T_u$ and $L_u$ are calculated from their local values.

One-sided velocity spectra $S_{uu}(f)$ were calculated after subtracting the complete-record mean.
Full-record periodograms resolve the peaks used in the periodic frequency comparison; the frequency spacing is 0.0333 Hz for a 30 s record.
Broadband spectral curves and downstream spectral comparisons use Welch estimates with 10 s Hann windows and 50\% overlap, giving a 0.1 Hz frequency spacing.
Cumulative variance fractions and band variances are calculated from full-record periodograms.
Each cumulative fraction is normalized by the total spectral variance of the corresponding record and measurement height.

For the downstream energy-share maps, each Welch spectral bin is normalized by the complete 0--Nyquist spectral sum at that height, including the zero-frequency bin.
The difference between shares at 4 and 2 m is expressed in percentage points (pp). This normalization describes frequency composition separately from the full-record velocity variance.

The velocity autocorrelation is evaluated using the noncircular, biased covariance normalization,
\begin{equation}
\rho_{uu}(\tau_\ell)=
\frac{\sum_{n=1}^{N-\ell}u'_n u'_{n+\ell}}
{\sum_{n=1}^{N}(u'_n)^2},\qquad \tau_\ell=\ell/f_s.
\label{eq:acf}
\end{equation}
where $N$ is the number of samples and $\ell$ is the lag in samples.
The integral time and streamwise integral length scale are
\begin{equation}
T_u=\int_0^{\tau_{0,u}}\rho_{uu}(\tau)\,\mathrm{d}\tau,
\qquad L_u=U T_u,
\label{eq:scale}
\end{equation}
where $\tau_{0,u}$ is the lag at which $\rho_{uu}$ first crosses zero, determined by linear interpolation.
The integral is evaluated numerically using trapezoidal integration.
The conversion to length uses the local mean velocity and Taylor's convection assumption.
For a periodically forced signal, $L_u$ also reflects the organized motion; its interpretation is therefore tied to the accompanying spectrum.

\subsection{Vane excitation and test conditions}
\label{subsec:excitation}
Three vane conditions were compared: stationary horizontal vanes, sinusoidal motion and pseudorandom motion with a prescribed spectrum.
For sinusoidal motion, the commanded angle of vane $j$ was
\begin{equation}
\theta_j(t)=A_p\sin(2\pi f_0t),\qquad j=1,\ldots,15,
\label{eq:periodic}
\end{equation}
where $A_p$ is the peak angle and $f_0$ is the forcing frequency.
The periodic reference is $20^\circ$ peak at 1 Hz.

Pseudorandom angle commands were constructed as a sum of harmonics with random phases using spectral representation,\cite{shinozuka_spectral_1991}
\begin{equation}
\theta_j(t)=A_r\sum_{k=2}^{150}\sqrt{2w_k}\cos(2\pi f_k t+\phi_k),
\qquad f_k=\frac{k}{T_c},\quad \sum_k w_k=1,
\label{eq:random}
\end{equation}
where $A_r$ is the RMS angle, $T_c=30$ s is the command period, and $w_k$ is the normalized variance weight at frequency $f_k$.
The phases $\phi_k$ were drawn uniformly from $[0,2\pi)$, and the same set was used for all three prescribed spectra.
The discrete frequencies span 0.0667--5 Hz. The upper limit was selected to protect the mechanical system, and the command was updated at 25 Hz.
The reference pseudorandom amplitude was $15^\circ$ RMS, close to the angular RMS of the periodic reference, $20^\circ/\sqrt{2}=14.14^\circ$.

\begin{figure*}[!htbp]
\centering
\includegraphics[width=\linewidth]{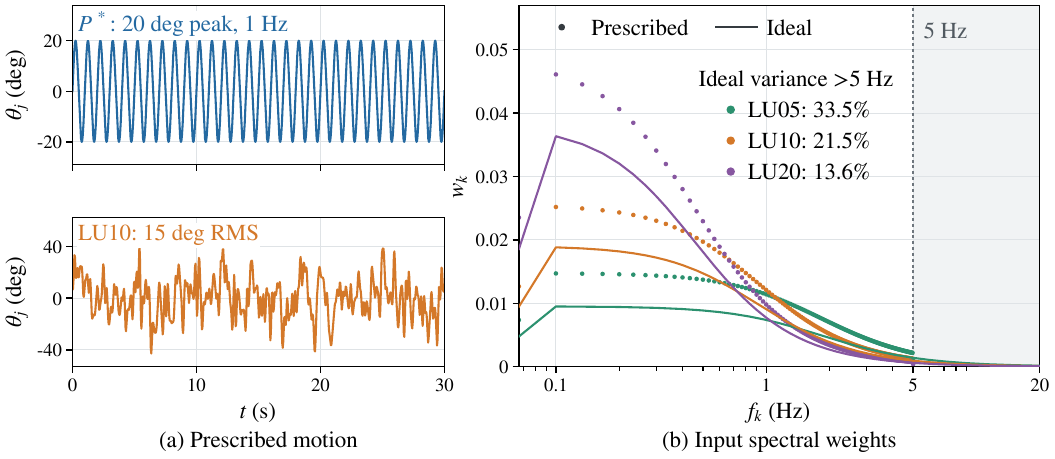}
\caption{Prescribed vane inputs: (a) periodic and LU10 command histories; (b) discrete variance weights normalized over the retained band (points) and the full ideal spectrum (lines). Shading marks frequencies above the 5 Hz cutoff; percentages give the ideal variance fraction from 5 Hz to infinity.}
\label{fig:inputs}
\end{figure*}

The weights $w_k$ were obtained by integrating a von K\'arm\'an-type spectrum over discrete frequency intervals and normalizing the resulting contributions,\cite{vonkarman_progress_1948}
\begin{equation}
G(f;\lambda)=\frac{4\lambda/U_d}{[1+70.8(f\lambda/U_d)^2]^{5/6}},
\qquad
w_k=\frac{\int_{\mathcal B_k}G(f;\lambda)\,\mathrm{d}f}
{\sum_m\int_{\mathcal B_m}G(f;\lambda)\,\mathrm{d}f}.
\label{eq:weights}
\end{equation}
Each interval $\mathcal B_k$ is bounded by adjacent frequency midpoints and truncated at the retained band limits.
Here $\lambda$ is the input length parameter and $U_d=7$ m/s is the design velocity.
The values $\lambda=0.5$, 1 and 2 m define LU05, LU10 and LU20, respectively.
Increasing $\lambda/U_d$ shifts relative weight toward lower frequencies.
The finite command band excludes part of the ideal spectrum.
The retained weights are renormalized to preserve the prescribed angular RMS, as illustrated in Fig.~\ref{fig:inputs}(b).
These parameters specify the angle-command spectrum; the resulting velocity spectrum and integral length scale are determined from measurements.

The normalized command autocorrelation, averaged over a complete command period, and its integral time are
\begin{equation}
\rho_{\theta\theta}(\tau)=\sum_{k=2}^{150}w_k\cos(2\pi f_k\tau),
\qquad T_\theta=\int_0^{\tau_{0,\theta}}
\rho_{\theta\theta}(\tau)\,\mathrm{d}\tau.
\label{eq:command-time}
\end{equation}
where $\tau_{0,\theta}$ is the lag at which $\rho_{\theta\theta}$ first crosses zero.
Averaging over a complete period removes cross-frequency terms, while normalization by the angular variance cancels the factor $A_r^2$.
Thus, changing $A_r$ leaves the normalized command autocorrelation and $T_\theta$ unchanged.
The first-zero integral times are 0.099, 0.155 and 0.250 s for LU05, LU10 and LU20.
These input time scales account for the discrete, finite frequency band and provide a reference for comparison with the measured $T_u$.
Representative command histories are shown in Fig.~\ref{fig:inputs}(a).

\begin{table*}[!htbp]
\caption{Summary of test conditions.}
\label{tab:comparisons}

\setlength{\tabcolsep}{3pt}
\begin{ruledtabular}
\begin{tabular}{lclcc}
Test group & Fan PWM (\%) & Vane input\footnotemark[1] & $x$ (m) & $y/W$ \\
\hline
Fixed-vane baseline & 20, 40, 60, 80, 100 & Fixed & 2.6 & 0 \\
Periodic frequency & 60 & 0.5, 1, 2 Hz & 2.6 & 0 \\
Periodic amplitude & 60 & $10^\circ$, $20^\circ$, $30^\circ$, $40^\circ$ peak & 2.6 & 0 \\
Pseudorandom amplitude & 60 & \shortstack[l]{LU10; $5^\circ$, $10^\circ$,\\$15^\circ$, $20^\circ$ RMS} & 2.6 & 0 \\
Prescribed spectrum & 60 & LU05, LU10, LU20 & 2.6 & 0 \\
Mean-flow setting & 40, 60, 80 & Fixed, periodic, LU10 & 2.6 & 0 \\
Spanwise position & 60 & Fixed, periodic, LU10 & 2.6 & $0,\ \pm0.25$ \\
Downstream position & 60 & \shortstack[l]{Fixed, periodic,\\LU05, LU20} & 2.0, 2.6, 4.0 & 0 \\
\end{tabular}
\end{ruledtabular}
\footnotetext[1]{Unless specified, periodic inputs use a $20^\circ$ peak amplitude at 1 Hz, and pseudorandom inputs use a $15^\circ$ RMS amplitude.}
\end{table*}

Table~\ref{tab:comparisons} summarizes the test conditions.
Amplitude and frequency effects were evaluated at $x=2.6$ m and $y/W=0$.
Fixed-vane references were measured at the same fan duty cycle and location as the corresponding active-input cases.
\section{Results and discussion}
\label{sec:results}

\subsection{Baseline flow and periodic gust generation}
\label{subsec:baseline-periodic}
\begin{figure*}[!htbp]
\centering
\includegraphics[width=.85\linewidth]{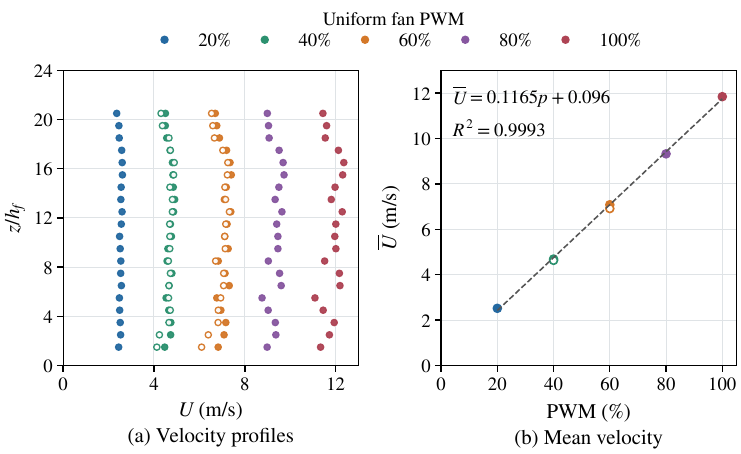}
\caption{Baseline flow with horizontal vanes at five uniform fan PWM settings: (a) mean-velocity profiles; (b) profile-mean velocity versus PWM. Open markers denote repeat measurements; the dashed line is a linear fit to the filled markers in (b).}
\label{fig:baseline}
\end{figure*}

Measurements with the vanes held horizontal establish the baseline velocity range and spatial uniformity.
Figure~\ref{fig:baseline} shows an approximately linear increase in profile-mean velocity with uniform fan PWM, from 2.52 m/s at 20\% to 11.84 m/s at 100\%.
A linear fit gives $\overline U=0.1165p+0.096$ m/s with $R^2=0.9993$, where $p$ is the PWM percentage value.
The fit uses one representative profile mean at each PWM setting; repeat measurements show the variation between records at the same setting.
This relationship provides a baseline for selecting the fan PWM for a desired mean velocity.

Uniform fan PWM produces nearly uniform mean-velocity profiles over the measured heights.
Table~\ref{tab:baseline} gives $CV_U$ values of 2.43--3.36\%, while the mean turbulence intensity remains between 2.91\% and 3.62\%.
At 60\% PWM, the selected fixed-vane reference has $\overline U=7.081$ m/s and $\overline{I_u}=3.60\%$.
These measurements provide the reference for assessing changes in mean velocity and fluctuation intensity under vane motion.

\begin{table}[!htbp]
\caption{Fixed-vane flow statistics over twenty measurement heights.}
\label{tab:baseline}
\begin{ruledtabular}
\begin{tabular}{cD{.}{.}{2.3}D{.}{.}{2.2}D{.}{.}{2.2}}
Fan PWM (\%) & \multicolumn{1}{c}{$\overline U$ (m/s)} & \multicolumn{1}{c}{$CV_U$ (\%)} & \multicolumn{1}{c}{$\overline{I_u}$ (\%)} \\
\hline
20 & 2.521 & 2.43 & 2.91 \\
40 & 4.661 & 3.36 & 3.34 \\
60 & 7.081 & 3.14 & 3.60 \\
80 & 9.322 & 2.99 & 3.62 \\
100 & 11.840 & 3.06 & 3.60 \\
\end{tabular}
\end{ruledtabular}
\end{table}

\begin{figure*}[!htbp]
\centering
\includegraphics[width=.93\linewidth]{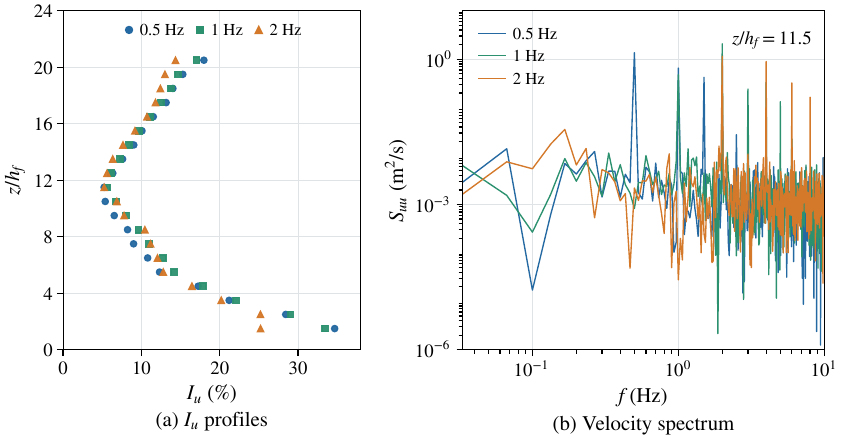}
\caption{Periodic frequency response at a $20^\circ$ peak angle and 60\% PWM: (a) turbulence-intensity profiles and (b) representative velocity spectra at $z/h_f=11.5$.}
\label{fig:periodic-frequency}
\end{figure*}

Periodic vane motion generates gusts with a controllable dominant frequency.
At a $20^\circ$ peak angle, the 0.5, 1 and 2 Hz inputs produce distinct peaks in the velocity spectra (Fig.~\ref{fig:periodic-frequency} and Table~\ref{tab:periodic-frequency}).
The dominant spectral peak occurs at the forcing frequency at most measurement heights, with second-harmonic dominance confined to a few points in the middle region.
As the forcing frequency increases, the profile-mean velocity integral time decreases but remains about 11\% of the forcing period ($f_0\overline{T_u}=0.105$--$0.112$).

\begin{table}[!htbp]
\caption{Dominant-peak frequencies and integral times under periodic forcing.}
\label{tab:periodic-frequency}
\begin{ruledtabular}
\begin{tabular}{lcccr}
Forcing frequency\footnotemark[1] & \multicolumn{2}{c}{Heights with dominant peak at} & \multicolumn{1}{c}{Mean integral time} & \multicolumn{1}{c}{\shortstack{Normalized\\integral time}} \\
$f_0$ (Hz) & $f_0$ & $2f_0$ & \multicolumn{1}{c}{$\overline{T_u}$ (s)} & \multicolumn{1}{c}{$f_0\overline{T_u}$} \\
\hline
0.5 & 18 & 2 & 0.224 & 0.112 \\
1 & 19 & 1 & 0.110 & 0.110 \\
2 & 20 & 0 & 0.052 & 0.105 \\
\end{tabular}
\end{ruledtabular}
\footnotetext[1]{Peak angular amplitude: $20^\circ$; fan PWM: 60\%. Counts refer to twenty measurement heights.}
\end{table}

The intensity profiles have a common low-response region near the middle of the sampled height range.
Their minima occur at $z/h_f=11.5$, where $I_u$ is approximately 5.23\%, 5.59\% and 5.24\% for the three frequencies.
For the 1 Hz input, the second harmonic dominates the local spectrum at this height; however, the intensity minimum also occurs under inputs for which the forcing frequency remains dominant there.
The recurring intensity minimum therefore cannot be attributed solely to second-harmonic dominance.

\begin{figure*}[!htbp]
\centering
\includegraphics[width=\linewidth]{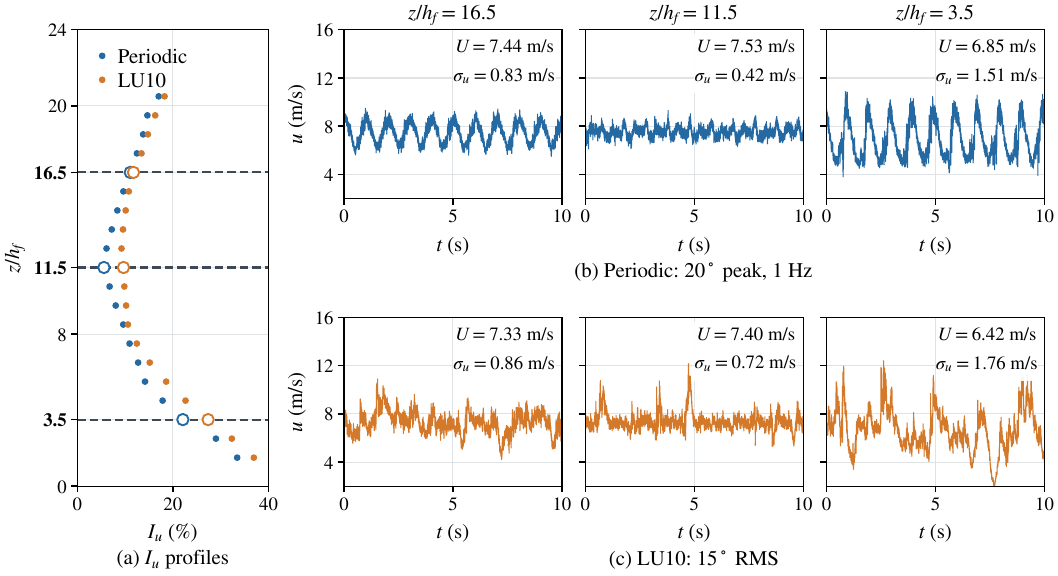}
\caption{Flow response to periodic and LU10 vane inputs at 60\% PWM: (a) turbulence-intensity profiles; (b) periodic and (c) LU10 velocity histories at the three heights marked in (a). The histories show 10 s excerpts; annotated $U$ and $\sigma_u$ are calculated from the full records.}
\label{fig:histories}
\end{figure*}

The velocity histories in Fig.~\ref{fig:histories}(b) help distinguish changes in fluctuation amplitude from changes in mean velocity.
At $z/h_f=16.5$ and 11.5, the mean velocities are 7.442 and 7.535 m/s, respectively, whereas $\sigma_u$ decreases from 0.826 to 0.421 m/s.
The smaller $I_u$ at the middle point therefore reflects weaker velocity fluctuations at nearly the same mean velocity.
The larger fluctuations at the lower representative point further show that identical vane commands produce different fluctuation amplitudes across the measured heights.

\begin{figure*}[!htbp]
\centering
\includegraphics[width=.75\linewidth]{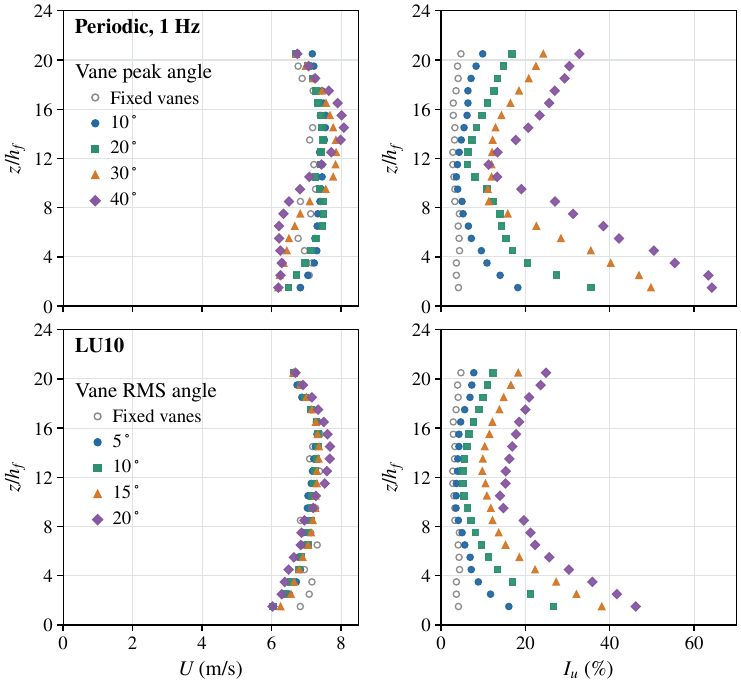}
\caption{Mean-velocity and turbulence-intensity profiles at 60\% PWM under increasing angular amplitudes. The upper row shows periodic forcing at 1 Hz, with peak angles in the legend; the lower row shows LU10 forcing, with RMS angles. Open gray markers denote the fixed-vane reference.}
\label{fig:amplitude}
\end{figure*}

Increasing the angular amplitude raises the periodic fluctuation intensity (Fig.~\ref{fig:amplitude}, upper row).
At 60\% PWM and 1 Hz, increasing the peak angle from $10^\circ$ to $40^\circ$ raises $\overline{I_u}$ from 7.33\% to 31.81\%.
Over the same scan, $\overline U$ decreases from 7.333 to 7.002 m/s, while $CV_U$ increases from 2.46\% to 10.08\% (Table~\ref{tab:amplitude}).
Larger angular excursions therefore produce stronger gusts together with a less uniform mean-velocity profile.

\begin{table*}[!htbp]
\caption{Velocity-profile statistics at 60\% fan PWM. Angular amplitudes are peak values for periodic forcing at 1 Hz and RMS values for LU10.}
\label{tab:amplitude}
\begin{ruledtabular}
\begin{tabular}{lcD{.}{.}{1.3}D{.}{.}{2.2}D{.}{.}{2.2}D{.}{.}{1.3}}
Input & \shortstack{Angular amplitude\\(deg)} & \multicolumn{1}{c}{$\overline U$ (m/s)} & \multicolumn{1}{c}{$CV_U$ (\%)} & \multicolumn{1}{c}{$\overline{I_u}$ (\%)} & \multicolumn{1}{c}{$\overline{\sigma_u}$ (m/s)} \\
\hline
Fixed & 0 & 7.081 & 3.14 & 3.60 & 0.254 \\[3pt]
Periodic & 10 & 7.333 & 2.46 & 7.33 & 0.532 \\
Periodic & 20 & 7.222 & 4.08 & 14.10 & 1.000 \\
Periodic & 30 & 7.128 & 8.43 & 21.98 & 1.506 \\
Periodic & 40 & 7.002 & 10.08 & 31.81 & 2.149 \\[3pt]
LU10 & 5 & 6.927 & 4.42 & 6.23 & 0.423 \\
LU10 & 10 & 6.977 & 4.98 & 10.24 & 0.696 \\
LU10 & 15 & 7.032 & 4.51 & 16.49 & 1.137 \\
LU10 & 20 & 7.033 & 7.14 & 23.04 & 1.582 \\
\end{tabular}
\end{ruledtabular}
\end{table*}

\subsection{Amplitude effects under pseudorandom forcing}
\label{subsec:pseudorandom}
At similar angular RMS values, replacing the sinusoidal command with LU10 produces irregular velocity fluctuations in place of the predominantly periodic response (Fig.~\ref{fig:histories}(b,c)).
With the normalized LU10 spectrum and fan setting held fixed, increasing $A_r$ from $5^\circ$ to $20^\circ$ RMS raises the profile-mean turbulence intensity from 6.23\% to 23.04\% (Fig.~\ref{fig:amplitude}, lower row).
The increase extends across most measured heights, but its magnitude varies with position and the middle-region minimum remains.

The profile-mean velocity remains near 7 m/s, spanning only 0.105 m/s across the LU10 scan (Table~\ref{tab:amplitude}).
However, $CV_U$ is 4.42\% at $5^\circ$ and 7.14\% at $20^\circ$, with changes in the local velocity profiles.
Thus, increasing the angular amplitude changes the mean-velocity distribution even though the profile-mean velocity remains nearly unchanged.

\begin{figure*}[!htbp]
\centering
\includegraphics[width=.85\linewidth]{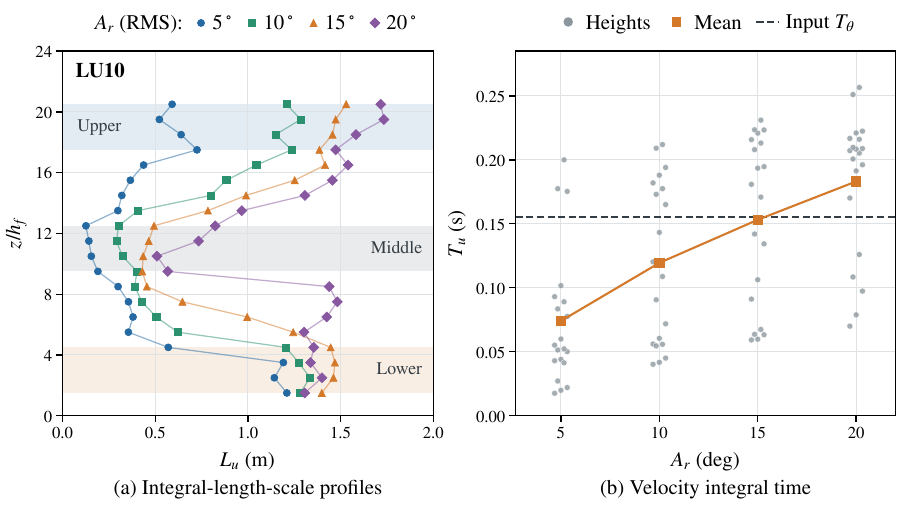}
\caption{Effect of LU10 angular RMS at 60\% PWM: (a) integral-length-scale profiles; (b) velocity integral times at twenty heights (circles) and their mean (squares). The dashed line marks the command integral time, $T_\theta=0.155$ s. Shading in (a) identifies the sampling regions used in Sec.~\ref{subsec:middle-deficit}.}
\label{fig:amplitude-time}
\end{figure*}

With the normalized LU10 command spectrum held fixed, increasing $A_r$ from $5^\circ$ to $20^\circ$ raises the profile-mean integral length scale from about 0.50 to 1.27 m (Fig.~\ref{fig:amplitude-time}(a)).
Over the same range, the profile-mean integral time increases from 0.074 to 0.183 s (Fig.~\ref{fig:amplitude-time}(b)), while the profile-mean velocity remains nearly unchanged.
The scale increase is therefore accompanied by longer velocity correlation times, consistent with the local relation $L_u=UT_u$.
This increase is spatially nonuniform: the middle-region minimum persists, and the lowest sampled heights show no consistent monotonic increase.

\begin{figure*}[!htbp]
\centering
\includegraphics[width=.8\linewidth]{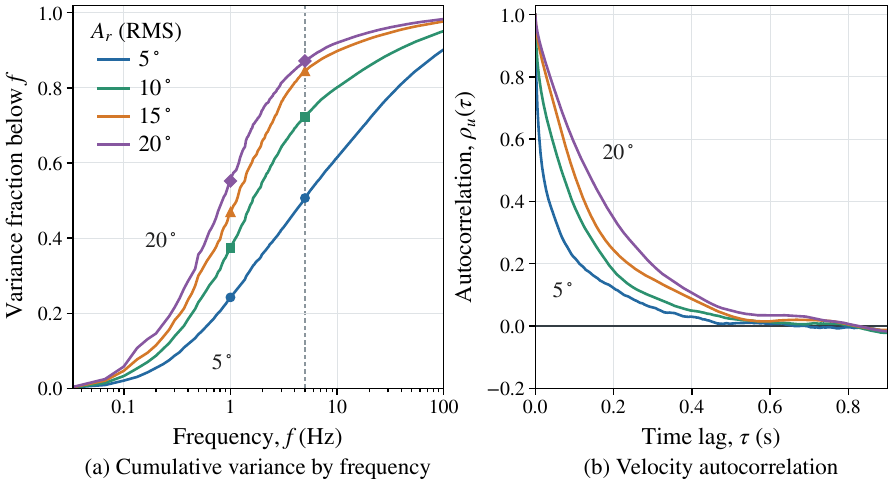}
\caption{LU10 amplitude effects at 60\% PWM: (a) cumulative variance fractions from full-record periodograms; (b) normalized velocity autocorrelations. Both quantities are normalized at each height before averaging over twenty heights.}
\label{fig:amplitude-spectral}
\end{figure*}

The measured variance becomes more concentrated at low frequencies as the amplitude increases (Fig.~\ref{fig:amplitude-spectral}(a)).
From $5^\circ$ to $20^\circ$, the height-averaged variance fraction below 1 Hz increases from 24.2\% to 55.2\%, while that below 5 Hz increases from 50.7\% to 87.2\%.
The normalized autocorrelations also decay more slowly at larger amplitudes (Fig.~\ref{fig:amplitude-spectral}(b)).
A uniform amplification of all frequency components would leave the normalized autocorrelation unchanged; the measured increase in $T_u$ instead accompanies a change in frequency composition, with greater weight at low frequencies.

One possible explanation is that slow fluctuations associated with vane motion become more prominent relative to background fluctuations as the angular amplitude increases.
The growing variance fraction below the 5 Hz command cutoff is consistent with this interpretation.
However, the measured spectra do not separate background and vane-related contributions, so this fraction cannot be interpreted as the fraction of variance generated by vane motion.

The measured integral time does not simply match the command integral time, $T_\theta=0.155$ s (Fig.~\ref{fig:amplitude-time}(b)).
At $15^\circ$, the mean $T_u/T_\theta$ is close to 1, but local ratios range from approximately 0.38 to 1.49; at $20^\circ$, the mean ratio reaches about 1.18.
Thus, even with the LU10 command spectrum held fixed, the measured velocity integral time varies with angular amplitude and height.

\subsection{Effects of the prescribed input spectrum}
\label{subsec:input-spectrum}
\begin{figure*}[!htbp]
\centering
\includegraphics[width=0.95\linewidth]{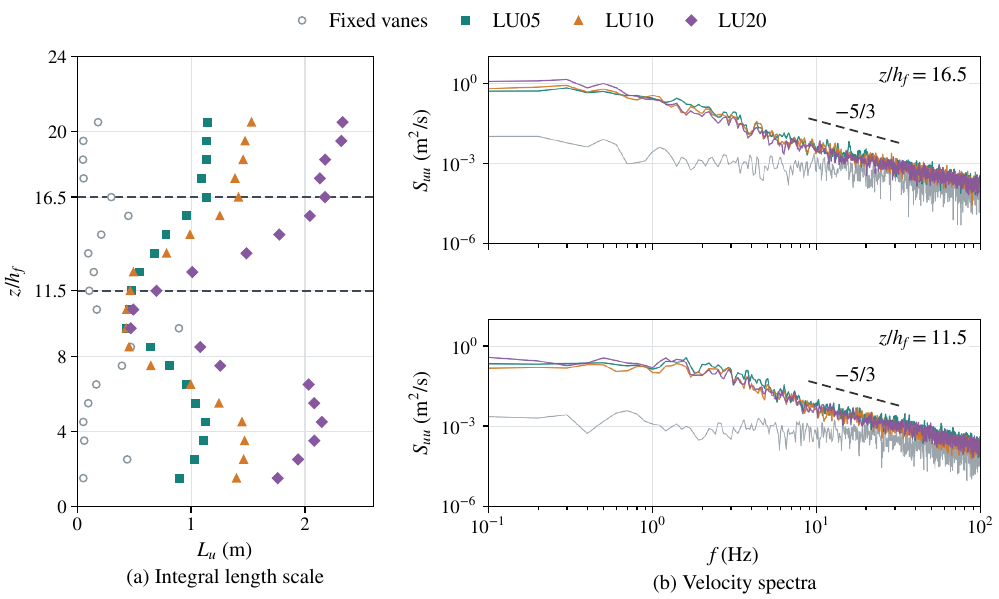}
\caption{Integral-length-scale profiles and velocity spectra under fixed and pseudorandom vane inputs at 60\% PWM. Horizontal guides mark the spectral measurement heights.}
\label{fig:scales-spectra}
\end{figure*}

At 60\% PWM and a common angular RMS of $15^\circ$, LU05, LU10 and LU20 produce profile-mean integral length scales of approximately 0.879, 1.062 and 1.674 m, respectively (Fig.~\ref{fig:scales-spectra}(a)).
The mean scale nearly doubles between LU05 and LU20, while the profile-mean turbulence intensities span 16.49\%--18.28\%.
Changing the prescribed spectrum therefore allows different mean scales to be obtained at similar, though not identical, turbulence intensities.

At both heights shown in Fig.~\ref{fig:scales-spectra}(b), pseudorandom forcing raises the spectral energy above the fixed-vane level over a broad frequency range, with the largest separation at low frequencies.
The fixed-vane references also have much shorter mean integral scales, approximately 0.1--0.2 m.

The differences between the prescribed spectra are less pronounced in the middle region.
Although LU20 produces a much larger profile-mean scale than LU05, their minimum $L_u$ values are approximately 0.47 and 0.43 m, respectively.
Thus, the separation in mean scale does not represent a comparable separation throughout the measured profile.

Together with the amplitude scan, these results show that increasing LU10 amplitude raises both mean intensity and mean scale, whereas changing the prescribed spectrum at fixed angular RMS adjusts the mean scale with a smaller variation in intensity.
This distinction is relevant when selecting inflows for tests sensitive to both quantities, including roof-pressure and heat-transfer studies.\cite{mokhtar_investigating_2025,sak_role_2007}

\subsection{Mean-velocity effects and spatial development}
\label{subsec:spatial}
\begin{figure*}[!htbp]
\centering
\includegraphics[width=.75\linewidth]{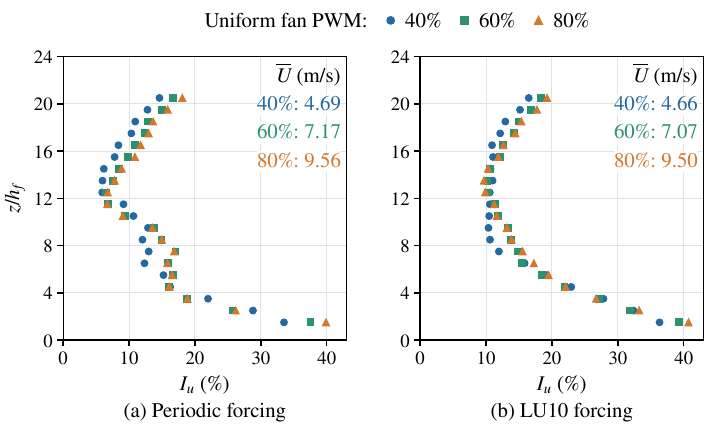}
\caption{Turbulence-intensity profiles at three uniform fan PWM settings under unchanged periodic or LU10 vane excitation. Insets give the corresponding profile-mean velocities.}
\label{fig:pwm}
\end{figure*}

With the vane command held fixed for each excitation, increasing PWM from 40\% to 80\% approximately doubles the profile-mean velocity from 4.7 to 9.5 m/s (Fig.~\ref{fig:pwm}).
The turbulence-intensity profiles retain broadly similar shapes, and the mean turbulence intensity increases from 13.44\% to 15.06\% under periodic forcing and from 15.97\% to 17.30\% under LU10 forcing.
These changes in turbulence intensity are modest compared with the approximately twofold increase in mean velocity.
Where the local $I_u$ remains similar, the relation $\sigma_u=UI_u$ implies an approximately proportional increase in absolute fluctuation amplitude with local mean velocity.

\begin{figure}[!htbp]
\centering
\includegraphics[width=.5\linewidth]{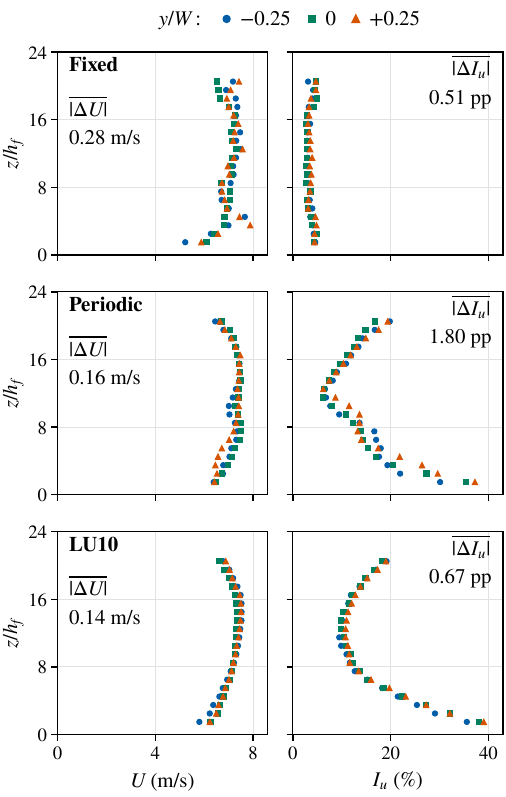}
\caption{Mean-velocity and turbulence-intensity profiles at three spanwise positions under fixed, periodic and LU10 vane inputs (top to bottom) at 60\% PWM and $x=2.6$ m. Insets give the mean absolute difference from the centerline, pooled over the two off-center profiles at twenty matched heights.}
\label{fig:spanwise}
\end{figure}

At $x=2.6$ m, the mean-velocity profiles at $y/W=-0.25$, 0 and $+0.25$ are similar under fixed, periodic and LU10 vane inputs (Fig.~\ref{fig:spanwise}).
Mean absolute velocity differences from the centerline range from 0.14 to 0.28 m/s.
The corresponding mean absolute difference in turbulence intensity is largest under periodic forcing (1.80 pp), compared with less than 0.7 pp for fixed vanes and LU10.
These differences are of the same order as the paired repeat-measurement differences, which include variation in both operation and measurement (Appendix~\ref{app:repeatability}, Table~\ref{tab:repeatability}).
Both active inputs retain the middle-region minimum and higher intensities toward the ends of the sampled profiles.

\begin{figure*}[!htbp]
\centering
\includegraphics[width=\linewidth]{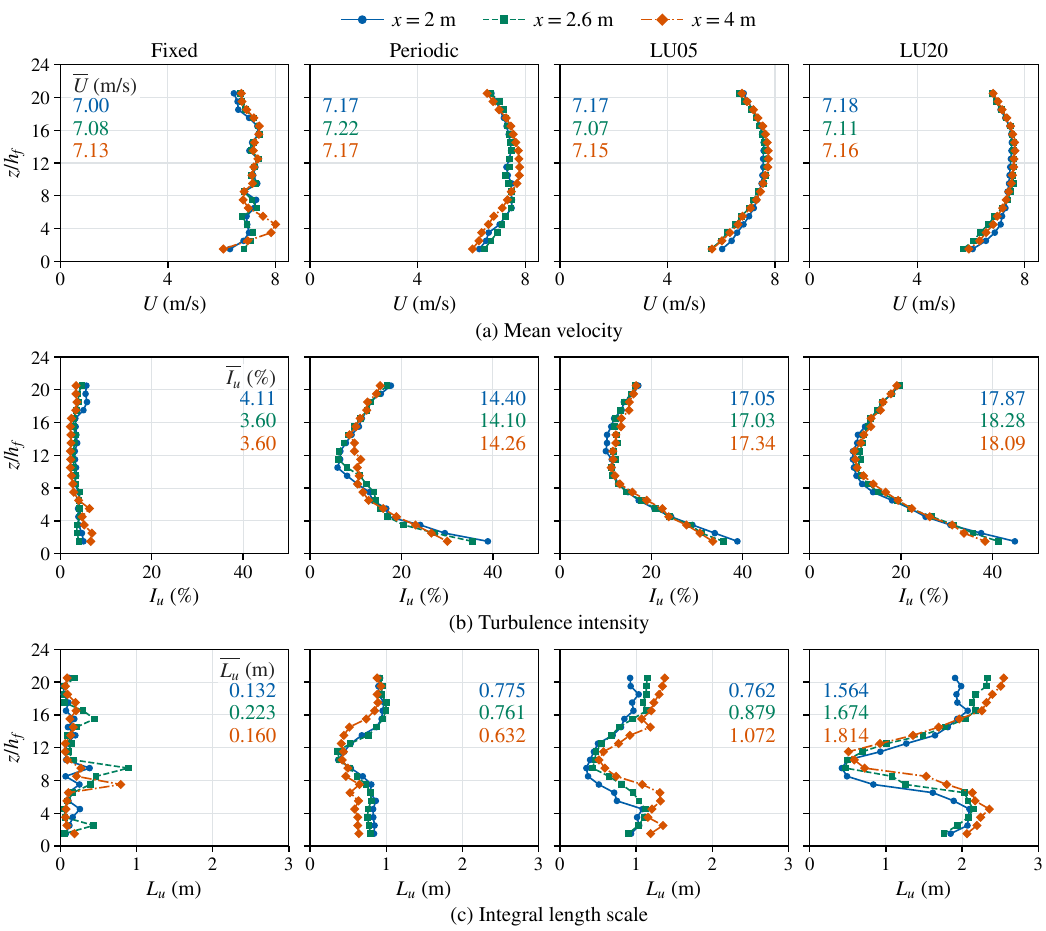}
\caption{Downstream profiles under fixed, periodic, LU05 and LU20 vane inputs: (a) mean velocity, (b) turbulence intensity and (c) integral length scale. Curves compare $x=2$, 2.6 and 4 m. Colored values give arithmetic means over the twenty measured heights.}
\label{fig:downstream}
\end{figure*}

Between $x=2.0$ and 4.0 m, the profile-mean velocities under active vane forcing remain near 7 m/s, and the turbulence-intensity profiles retain their overall shapes (Fig.~\ref{fig:downstream}).
For LU05 and LU20, the mean intensities change little, while the mean integral scales increase from 0.762 to 1.072 m and from 1.564 to 1.814 m, respectively.
The scale at 4 m exceeds that at 2 m at all measured heights for LU05 and at most heights for LU20.
LU20 retains the larger mean scale at all three stations.

\begin{figure*}[!htbp]
\centering
\includegraphics[width=\linewidth]{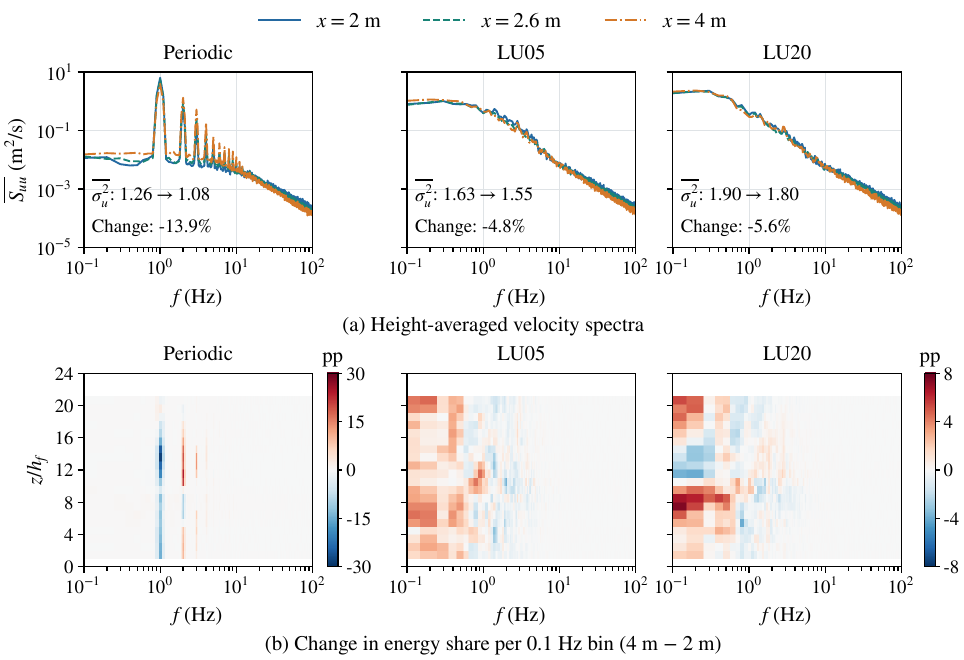}
\caption{Downstream spectral evolution: (a) velocity spectra averaged over twenty heights; (b) changes in local variance fractions per 0.1 Hz band (4 m minus 2 m, pp). Annotations in (a) give mean full-record variances (m$^2$/s$^2$) at 2 and 4 m and their percentage changes. Fractions in (b) use the total spectral variance up to the Nyquist frequency.}
\label{fig:downstream-spectra}
\end{figure*}

At all three stations, the height-averaged spectra retain distinct peaks under periodic forcing and broadband distributions under LU05 and LU20 (Fig.~\ref{fig:downstream-spectra}(a)).
From 2 to 4 m, the height-averaged velocity variance decreases by only about 5\% for both LU05 and LU20, while their mean integral length scales increase.

Under periodic forcing, the height-averaged spectrum has a weaker fundamental peak at 1 Hz and a stronger second-harmonic peak at 2 Hz at $x=4$ m than at $x=2$ m.
The ratio of spectral density at 2 Hz to that at 1 Hz increases from about 0.10 to 0.34.
Each trend is observed at most measurement heights.
Over the same distance, the profile-mean $L_u$ decreases from 0.775 to 0.632 m.
The greater relative prominence of the second harmonic may contribute to the shorter correlation time, although the full spectral distribution, rather than the two peak values alone, governs the autocorrelation.
Here, $L_u$ reflects both periodic and irregular fluctuations, so its decrease does not by itself imply smaller turbulent eddies.

The principal spectral features persist downstream, but local frequency composition (Fig.~\ref{fig:downstream-spectra}(b)) and integral scale continue to change.
Inflow statistics should therefore be characterized at the intended test station.

\subsection{Middle-region deficits in turbulence intensity and integral scale}
\label{subsec:middle-deficit}
The middle-region minimum in turbulence intensity persists across changes in forcing frequency, amplitude and waveform (Figs.~\ref{fig:periodic-frequency}--\ref{fig:amplitude}).
The pseudorandom cases also exhibit a minimum in $L_u$ in this region, although the two minima need not occur at the same height (Fig.~\ref{fig:scales-spectra}).
In the representative LU10 histories (Fig.~\ref{fig:histories}(c)), the upper- and middle-point mean velocities differ by about 0.9\%, while the middle-point $\sigma_u$ is about 17\% lower.
At these points, the intensity difference therefore mainly reflects weaker velocity fluctuations.

To examine the frequency dependence of this contrast, four measured heights are grouped within each of three ranges: $z/h_f=17.5$--$20.5$ (upper), $9.5$--$12.5$ (middle), and $1.5$--$4.5$ (lower), shaded in Fig.~\ref{fig:amplitude-time}(a).
Band variances are calculated from full-record periodograms at each height and then averaged within each region.

\begin{figure*}[!htbp]
\centering
\includegraphics[width=\linewidth]{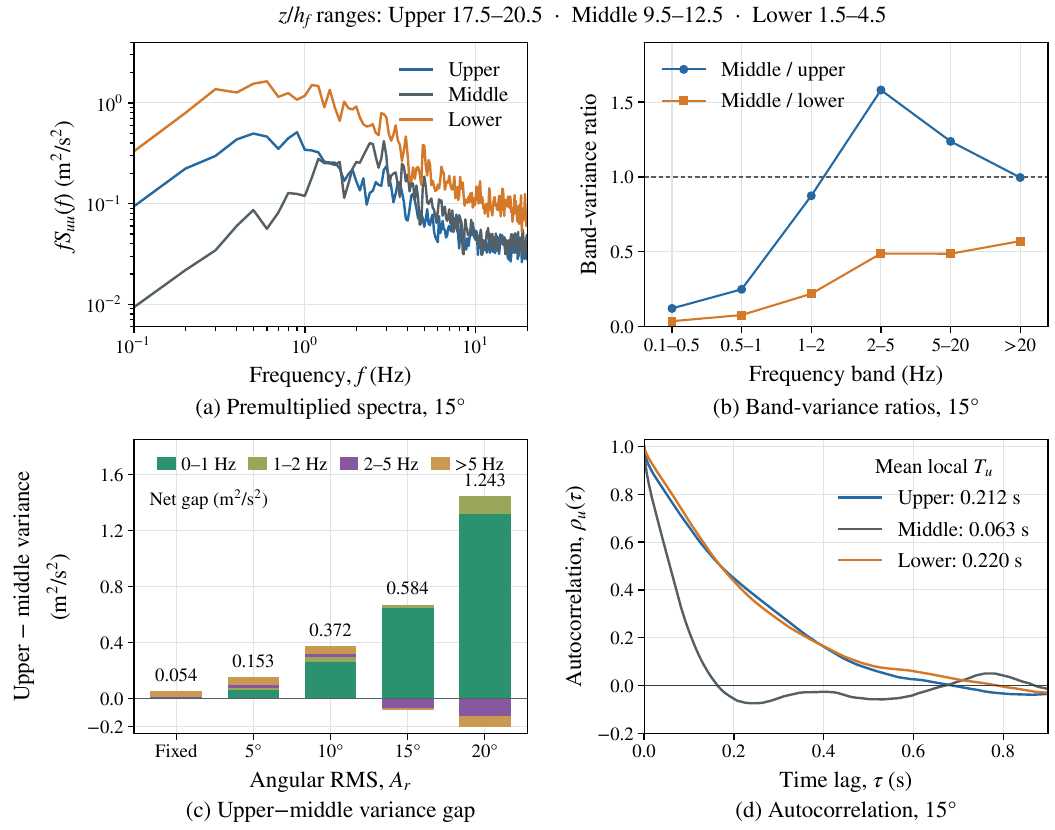}
\caption{Middle-region deficits at 60\% PWM, $x=2.6$ m and $y=0$: (a) premultiplied spectra; (b) band-variance ratios; (c) signed band-variance differences (upper minus middle); (d) regional means of normalized local autocorrelations. Panels (a,b,d) use LU10 at $15^\circ$ RMS. Bar annotations give net variance gaps (m$^2$/s$^2$).}
\label{fig:middle-spectral}
\end{figure*}

Under LU10 forcing at $15^\circ$ RMS, the middle region contains less low-frequency variance than both the upper and lower regions (Fig.~\ref{fig:middle-spectral}(a,b)).
The middle-to-upper band-variance ratio is 0.12 at 0.1--0.5 Hz but 1.58 at 2--5 Hz, whereas the middle-to-lower ratio remains below 1 in every displayed band.
Thus, the lower total variance in the middle region does not imply lower variance at every frequency: some bands contain more variance than in the upper region.

To compare the variance differences across amplitudes, let $V_{r,k}$ denote the mean variance in region $r$ and frequency band $k$.
The upper-minus-middle differences in Fig.~\ref{fig:middle-spectral}(c) are
\begin{equation}
\Delta V_k=V_{\mathrm{upper},k}-V_{\mathrm{middle},k},
\qquad \Delta V=\sum_k\Delta V_k,
\label{eq:regional-variance-gap}
\end{equation}
where the sum covers the nonoverlapping bands spanning the full spectrum.

The total difference, $\Delta V$, increases from 0.054 m$^2$/s$^2$ with fixed vanes to 1.243 m$^2$/s$^2$ under LU10 forcing at $20^\circ$ RMS.
The increase is concentrated below 1 Hz, whereas the small fixed-vane difference arises mainly above 5 Hz.
At $15^\circ$ and $20^\circ$, negative $\Delta V_k$ values in the 2--5 Hz and above-5 Hz bands indicate greater middle-region variance, partly offsetting the positive low-frequency difference.

The middle-region minimum in $L_u$ is accompanied by a shorter correlation time.
For the $15^\circ$ LU10 case, the regional means of the pointwise $T_u$ are 0.212, 0.063 and 0.220 s in the upper, middle and lower regions, respectively (Fig.~\ref{fig:middle-spectral}(d)).
Lower variance alone cannot explain this difference because the autocorrelations are normalized.
The faster middle-region decay is consistent with the smaller relative contribution of low-frequency fluctuations; through $L_u=UT_u$, the shorter correlation time contributes to the length-scale deficit.

The regional comparisons show that the response under common vane forcing varies with both height and frequency.
Changes in blockage and wake position can alter local longitudinal responses in moving-grid and vane experiments.\cite{wester_how_2022,french_design_2021}
In the present facility, height-dependent vane chord and tunnel confinement may contribute to the spatial variation in the low-frequency response, but their roles in the middle-region deficit remain unresolved.

For a specific experiment, the turbulence-intensity and integral-scale profiles should be evaluated over the intended test region against the required flow conditions. 
The measured amplitude and spectral responses provide a basis for selecting vane inputs, 
while the persistent middle-region deficits motivate exploring height-dependent actuation to improve local flow conditions.
\section{Conclusions}
\label{sec:conclusions}
Controlling inflow turbulence intensity and integral scale enables systematic investigation of unsteady-flow effects in wind-tunnel experiments.
Prescribed-spectrum pseudorandom vane excitation generated broadband turbulence with large integral scales, whereas periodic vane motion generated gusts at 0.5--2 Hz.
The experiments quantified how turbulence intensity and integral scale varied with vane angular amplitude and command spectrum.
The main findings are as follows.

\begin{enumerate}
\item Increasing angular amplitude raises the fluctuation intensity under both periodic and pseudorandom forcing.
At 60\% PWM, the tested amplitude ranges produced profile-mean turbulence intensities of 7.3\%--31.8\% under periodic forcing and 6.2\%--23.0\% under LU10 forcing.
The profile-mean velocity changed comparatively little, but the mean-velocity profiles became less uniform at the larger amplitudes.

\item The measured integral scale depends on both angular amplitude and prescribed spectrum.
With the normalized LU10 input spectrum held fixed, increasing amplitude increased the relative contribution of low-frequency velocity fluctuations, accompanied by longer correlation times and larger integral scales.
At a fixed angular RMS of $15^\circ$, changing the prescribed spectrum produced profile-mean integral scales of 0.88--1.67 m at similar turbulence intensities.

\item For each of the two inputs (periodic and LU10), the same time-varying vane command was applied at all tested fan duty cycles from 40\% to 80\%. 
The turbulence-intensity profiles retained broadly similar shapes as the profile-mean velocity increased from about 4.7 to 9.5 m/s. 
The principal features of the height-averaged velocity spectra persisted between 2 and 4 m downstream. 
Over the same distance, the profile-mean integral length scale increased in the LU05 and LU20 cases and decreased in the periodic case.

\item Middle-region deficits in turbulence intensity and integral scale persisted under common vane commands.
For LU10 at $15^\circ$ RMS, the middle region had less low-frequency variance and faster normalized autocorrelation decay than the upper and lower regions.
At larger LU10 amplitudes, frequencies below 1 Hz dominated the upper-minus-middle variance difference.
These results associate the intensity deficit with weaker velocity fluctuations and the scale deficit with shorter correlation times, consistent with a smaller relative contribution of low-frequency fluctuations.
\end{enumerate}

The measured responses to vane angular amplitude and the prescribed command spectrum provide a basis for selecting vane inputs. 
The resulting turbulence intensity and integral length scale should be evaluated over the intended test region against the required inflow conditions. 
Further measurements of the local flow structure and additional velocity components could clarify the origin of the middle-region deficits, and tests with height-dependent vane commands could assess whether these deficits can be reduced.

\begin{acknowledgments}
This work was supported by the National Natural Science Foundation of China (Grant Nos. 52278493 and  W2541002) 
and the Guangdong Basic and Applied Basic Research Foundation (Grant No. 2024A1515240044).
The authors also thank Hao Hong, Meng Hu, and K. B. Rajasekara Babu for their assistance with the vane-array development and experiments.
\end{acknowledgments}

\section*{AUTHOR DECLARATIONS}

\subsection*{Conflict of Interest}
The authors have no conflicts to disclose.

\subsection*{Author Contributions}
Yutong Liu: Conceptualization; Data curation; Formal analysis; Investigation; Methodology; Software; Validation; Visualization; Writing -- original draft; Writing -- review \& editing.

Gang Hu: Conceptualization; Formal analysis; Funding acquisition; Methodology; Project administration; Resources; Supervision; Writing -- review \& editing.

Hongwei Zhang: Funding acquisition; Project administration; Supervision; Writing -- review \& editing.

\section*{DATA AVAILABILITY}
The data that support the findings of this study are available from the corresponding author upon reasonable request.

\appendix
\setcounter{table}{0}
\renewcommand{\thetable}{A\arabic{table}}
\section{Repeatability of measured flow statistics}
\label{app:repeatability}
Six settings were measured twice at the same twenty heights on the main centerline.
The active-input records lasted 30 s; each fixed-vane pair comprised one 20 s and one 30 s record.
For a statistic $X$, the mean difference and mean absolute error (MAE) between matched heights are
\begin{equation}
\Delta\overline X=\frac{1}{N_p}\sum_{i=1}^{N_p}(X_{2,i}-X_{1,i}),\qquad
\mathrm{MAE}_X=\frac{1}{N_p}\sum_{i=1}^{N_p}|X_{2,i}-X_{1,i}|.
\label{eq:paired-differences}
\end{equation}
The differences compare two measured profiles and characterize combined variation in operation and measurement.
For $U$ and $L_u$, percentage values use the paired profile mean, $(\overline X_1+\overline X_2)/2$, as the denominator. Differences in $I_u$ are expressed in pp.

\begin{table*}[!htbp]
\caption{Differences between paired measurements at matched heights. Signed differences are record 2 minus record 1; dashes indicate omitted relative differences for fixed vanes.}
\label{tab:repeatability}
\begin{ruledtabular}
\begin{tabular}{lD{.}{.}{2.2}D{.}{.}{1.2}D{.}{.}{2.2}D{.}{.}{1.2}cD{.}{.}{2.2}}
& \multicolumn{2}{c}{$U$ differences (\%)} & \multicolumn{2}{c}{$I_u$ differences (pp)} & \multicolumn{2}{c}{$L_u$} \\
Input & \multicolumn{1}{c}{\shortstack{Signed mean\\diff.}} & \multicolumn{1}{c}{MAE} & \multicolumn{1}{c}{\shortstack{Signed mean\\diff.}} & \multicolumn{1}{c}{MAE} & $\overline{L}_{u,1},\ \overline{L}_{u,2}$ (m) & \multicolumn{1}{c}{MAE (\%)} \\
\hline
Fixed, 40\% PWM & -1.28 & 2.42 & +0.05 & 0.57 & 0.134, 0.177 & \multicolumn{1}{c}{---} \\
Fixed, 60\% PWM & -2.42 & 2.74 & -0.06 & 0.48 & 0.223, 0.181 & \multicolumn{1}{c}{---} \\
Periodic & -1.33 & 3.02 & +1.11 & 2.09 & 0.797, 0.724 & 10.02 \\
LU05 & -0.15 & 3.31 & +1.64 & 1.68 & 0.819, 0.939 & 16.92 \\
LU10 & +1.03 & 3.11 & +0.96 & 1.11 & 1.114, 1.009 & 11.16 \\
LU20 & +1.34 & 4.52 & +2.35 & 2.35 & 1.649, 1.699 & 11.72 \\
\end{tabular}
\end{ruledtabular}
\end{table*}

For active inputs, the $U$ MAE is 3.0\%--4.5\%, whereas the mean difference remains within approximately $\pm1.4\%$ (Table~\ref{tab:repeatability}).
Thus, a small shift in the profile mean can coexist with larger local differences.
The corresponding MAE ranges are 1.1--2.35 pp for $I_u$ and 10\%--17\% for $L_u$.
Both measurements retain the profile-mean ordering LU05 $<$ LU10 $<$ LU20; LU20 also exceeds LU10 at all twenty matched heights in both records.

These paired differences provide a reference for interpreting the measured changes, but do not constitute confidence intervals or a complete measurement-uncertainty estimate.

\bibliography{references}

\end{document}